# A Complex Event Processing Approach for Crisis-Management Systems

*Exploiting Crowd Sourcing and Crowd Sensing Information for Situational Awareness*

Massimiliano L. Itria, Alessandro Daidone
ResilTech S.R.L.
Pontedera (Pisa), Italy

Andrea Ceccarelli
University of Florence
Florence, Italy

*Abstract*—In modern advanced emergency management systems many solutions for decision support have been provided as attempts to support humans to take important decisions for the critical situations recovery. The critical situation detection is a complex procedure that involves both human and machine activities and leads to take a decision for the management and situation recovery. This paper presents an approach for critical situation detection which uses event correlation technologies performing online analysis of real events through a Complex Event Processing architecture. Event correlation is used to relate events gathered from various sources, including crowd sensing and crowd sourcing sources, for detecting patterns and situations of interest in the emergency management context.

## I. Introduction

Online stream data processing has become a very important technology for many applications, such as social-media channels observation, activity recognition from video contents, computer network monitoring, trader behavior evaluation in financial markets or patient monitoring in some health facilities. In all of these applications, the amount of data being generated requires online processing and immediate reaction in order to be managed in an efficient way. Nowadays, the high availability of several type of sensors deployed on the territory suggests to exploit the online processing in many application domains such as surveillance and protection of critical infrastructures and areas, for example: train stations, airports, world heritage protected areas in some cities of art and so on.

The proliferation of modern mobile devices, such as smart-phones and tablets, has given a boost to the experimentation in the context of emergency management systems allowing the integration of the crowd sourcing [1] and crowd sensing [2] technologies. Crowd sourcing is the process of getting information online, from a crowd of people, while crowd sensing refers to the involvement of a large, diffuse group of participants in the task of retrieving reliable data from a specific field. By means of the possibility to easily link persons, facts, events and places through a large quantity of online geo-referenced data, users are the real holders of the "living information" and the producers of current information about social phenomena and dangerous events. Hence users may be considered as real "human sensors" providing qualitative, and sometimes quantitative, information.

The integration of information retrieved from mobile devices, from social media and from several type of sensors deployed in the infrastructures, allows the online analysis of a large amount of data used to detect and identify dangerous events. Such online approach enables the detection of critical situations as soon as they happen, so that a corresponding reaction can be successfully performed. In a nutshell, this mechanism aims to timely recognize (or even predict) critical situations, usually called Real-time Situational Awareness (RTSA, [15]). The main goal of RTSA is to recognize the critical situations in the given application domain as soon as possible in order to be able to take a decision for facing them properly. The decision is the first step of the reaction and it should be made by humans using a Decision Support System (DSS [16]) that helps them to decide how to face the emergency. The process, starting from the data extraction, leads to the detection of the situation in progress. It introduces several challenges: (i) first of all it should be highly efficient in order to handle a huge amount of data and detect the situation in progress before it is too late to perform the reaction successfully [11]; (ii) furthermore it should be able, when it is possible, to detect critical situations before they happen (early warning) in order to prepare a preventive action; (iii) it should be also tolerant to different types of noise, meaning that the process should acknowledge only trusted information from trusted sources, otherwise it could lead to wrong scenario definitions and consequently wrong decisions; and (iv) it should be sufficiently reliable to trust the logged events, including architecture resilience and trustworthy data collection [13], [14], possibly allowing forensic analysis [12].

Complex Event Processing (CEP, [3]) technology aims to resolve these challenges allowing an efficient management of the pattern detection process in the huge and dynamic data streams and as such it is very suitable for recognizing complex events and situations online.

This paper presents a CEP application RTSA using crowd sensing/sourcing technology in the context of the research project Secure! [21]. Furthermore it gives an overview of the Secure! Framework that is designed to detect critical situations managing input data from multimodal sources and providing decision support to the Secure! human operators. Afterwards, the paper describes the *Event Extraction and Integration* logical level, that has been developed in the context of the Secure! project, and the events correlation engine here located, that is able to recognize complex events using CEP technology. The approach used for designing the Secure! correlation engine focused on the following requirements: (i) the correlation module have to be adaptable to the possible changes of the environment in which the events happen; and (ii) the correlation has to consider also historical data in order to evaluate the actual events as the result of an off-line analysis on historical data mixed with online data.

The rest of the paper is organized as follows. Section II presents related works on crowd sensing and crowd sourcing *Information Fusion* technologies. Section III introduces fundamentals on CEP. Section IV presents the Secure! Framework logical architecture. Section V describes the *Event Processing and Management* component that has been have been developing and which is in charge of performing event correlation. Section VI presents insights on the *information fusion* approach adopted in the *Event Processing and Management* component. Section VII shows a case study in which a set of events is correlated, and finally Section VIII presents conclusion.

## II. RELATED WORK

Several works about crowd sensing and crowd sourcing *Information Fusion* technologies are available in literature. Most of them use crowd information for making market surveys. Some of them face the public protection problem for supporting police investigation as in [5], but the aim of that work does not comprise the RTSA for the emergency management. In [6] authors propose a novel information system working on mobile stations for data collection (about radioactivity and toxic material) and critical situation management due to pollution. In [8] problems of the European communication infrastructures (Tetra, GSM, Citizen Band, IP) are faced such as low interoperability and availability level; it is proposed a fast malfunction recovery exploiting network information correlation for malfunction detection. In [9] it is given an overview of the open research challenges in applying CEP for RTSA; the major weakness of [9] is that only the video content and social media observation is taken as input of the system, and mobile devices information are not considered.

## III. COMPLEX-EVENT PROCESSING TECHNOLOGY

This section introduces CEP, explaining the concept of event-driven behavior and defining the relevant terminology.

### A. Events and Situations

In [3], the term *event* is defined as "an occurrence within a particular system or domain; it is something that has happened, or is contemplated as having happened in that domain". This definition places the *event* concept into two different contexts: (i) the real world in which events happens and (ii) the realm of computerized event processing, where the word *event* is used to mean a programming entity that represents this occurrence. In the sphere of the Emergency Support Systems (ESS) are considered those events that happen in the real world and are represented in computing systems through information entities. Event processing allows to detect critical situation in order to response timely to the emergency. For the purpose of our work, that is managing events constituted by textual description of facts and involved entities (person or object), we define *micro-events* and *complex-events*. *Micro-events* belong to a basic event taxonomy [10] and represent simple real events involving only one entity for example: people detection, fire presence, impulsive sound recognition, object detection. On the other hand, *complex-events* are the aggregation result of the information contained in a set of *micro-events* which are correlated by spatial, temporal and causal relations defined by correlation rules. With the term *situation*, as defined in [4], we intend "one or more event occurrence that might require a reaction". When a critical situation happens a number of specific *complex-events* occur, the commixture of these *complex-events* identifies the specific *situation* in progress requiring an appropriate reaction: provide first aid, police action, recovery service. This is the reasoning behind the concept of event-driven behavior.

### B. Complex Event Processing

CEP consists of the processing of events generated by the combination of data from multiple sources and aggregated in *complex-events* representing situations or part of them [17], [18]. Common event processing operations include reading, creating, transforming, and deleting events. CEP is the means that allows to: extract relevant *micro-events* from several data streams belonging to event producers, correlate *micro-events* in data streams and aggregate information in *complex-events* reducing redundancy, computing complexity and uncertainty. We can consider CEP a means to achieve *information Fusion*.

## IV. THE SECURE! FRAMEWORK

The Secure! Project aims to provide ICT means and support services for the public and private security management exploiting the synergy between social media and crowd sourcing/sensing technologies. This section presents the Secure! Framework by describing its main features. To better understand the Secure! Framework, we describe the overall architecture of the reference system.

### A. The Secure! Framework

The Secure! Framework, currently in development phase, is a novel Decision Support System (DSS) for emergency management. It exploits information retrieved from a large quantity and several type of sensors deployed in the area of interest, in order to detect critical situations and perform the corresponding reaction. The Secure! Framework should also be able to detect critical situations before they happens analyzing *micro-events* provided by the social media and correlating them with: historical data and the *micro-events* from other sources. For example, threats to persons or things may be detected making a syntactic analysis of the text content provided by social media; searching for particular keywords, it is possible to recognize the intentions of a spiteful person.

The logical architecture of the Secure! Framework is depicted in Fig. 1. It is basically composed of four different logical levels, each of which comprises logical components and services. Input data comes from the following sources: (i) social media, (ii) web sites, (iii) mobile devices and embedded sensors (GPS, gyroscope, accelerometer, thermometer, proximity sensor), (iv) interactive television, (v) sensor networks in critical infrastructures. Starting from the bottom level in the diagram, data is received, collected, homogenized, correlated and aggregated in order to produce the *Secure! Situation* for the DSS system *Action Management and Decision Making* represented in the top level.

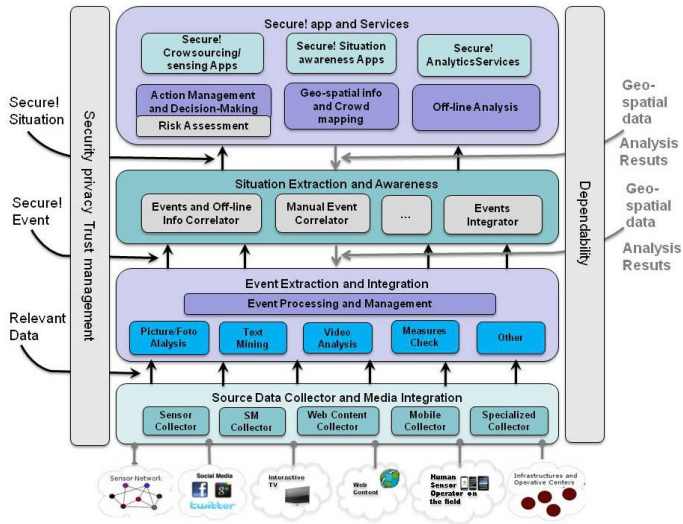

Fig. 1. Logical Architecture of Secure! Framework [21].

## V. THE EVENT PROCESSING LAYER

This section focuses on the online correlation component that we designed and developed relying on the CEP technology Esper [7]. It is integrated in the *Event Extraction and Integration* layer of the Secure! Framework and represents the fundamental topic of the work presented in this paper.

### A. Event Processing and Management Component

The *Event Processing and Management* (EPM) component aims to detect/recognize *micro-events*, classifying them depending on the basic event taxonomy established and producing *complex-events* through information fusion process. The logical architecture of the EPM component is shown in Fig. 2. Several modules in EPM cooperate for managing, storing, correlating and aggregating events. For brevity, this paper describes only the *Event Processor* module that is the core of the EPM component. The EPM receives *micro-events* from *Micro-event Producers,* designated to extract data from the sources, and executes the *micro-event* correlation. It can be configured by Secure! operators which can define or modify the correlation rules.

### B. Event Processor

The *Event processor* module is implemented relying on the *Esper* event correlation engine. The *Esper* engine works a bit like a database turned upside-down. Instead of storing the data and running queries against stored data, the *Esper* engine allows applications to store queries and run the data through. Response from the *Esper* engine is online when conditions occur that match queries. The execution model is thus continuous rather than only when a query is submitted. *Esper* offers event stream queries that provide the windows, aggregation, joining and analysis functions for use with streams of events. These queries are expressed through the Event Programming Language (EPL) [7] syntax. EPL has been designed for similarity with the SQL query language but differs from SQL in its use of views rather than tables. Views represent the different operations needed to structure data in an event stream and to derive data from an event stream. *Esper* is totally developed in Java, for this reason it has been integrated easily in the EPM component as a Java library. The Event Processor receives *micro-events* streams, applies the rules to them and returns the sets of *micro-events* that satisfy the rules.

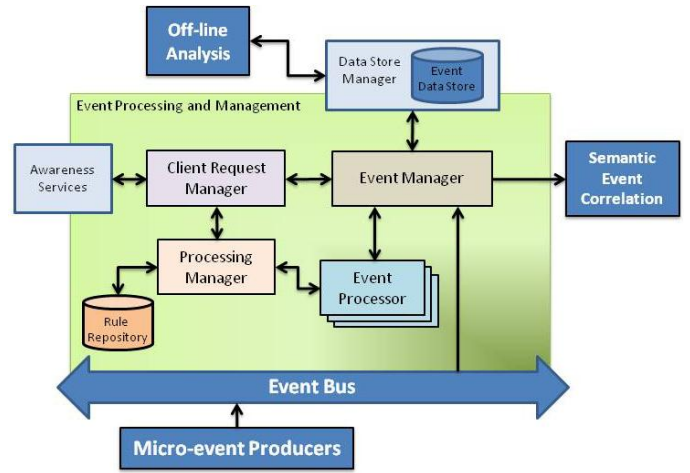

Fig. 2. Logical Architecture of the Event Processing and Management (EPM) component.

## VI. INFORMATION FUSION

*Information fusion* consists of merging of information from heterogeneous sources with differing conceptual, contextual and typographical representations, so as to answer questions of interest and take proper decision. It involves the combination of information into a new set of information towards reducing uncertainty. In our work, *information fusion* is applied to the *micro-events* in input to the EPM component in order to produce *complex-events* for the successive phase of situation detection. Among other things, the *micro-events* contains the texture description of the real event, the time when it happened, the entity involved and the source that generated it.

### A. Event Information Fusion

The EPM achieves the *Information Fusion* in three fundamental phases: (i) *Syntactic Check and Priority Allocation*, (ii) *Event Merging*, (iii) *Event Trust Analysis*. The first phase consists of the *micro-events* coherence check verifying their content and searching for relevant keywords in the specific application domain. Analyzing the found keywords, a priority value used by the system for managing emergency situations before others is assigned to the *micro-event*. As soon as the *micro-events* are syntactically correct, they are ready to be correlated and merged in a *complex-event.*

In the *Event Merging* phase EPM applies the correlation rule sets to them and aggregates the correlated *micro-events* in the *complex event*. The *complex-event* contains more detailed and complete information than *micro-event*, it suggests a situation in progress or a part of it.

The last phase, *Event Trust Analysis*, is an essential process useful to check the *complex-event* trustworthiness tracing the sources. The *Secure! Event* is defined as a trusted *complex-event.* Fig. 3 shows the *Event Fusion* process.

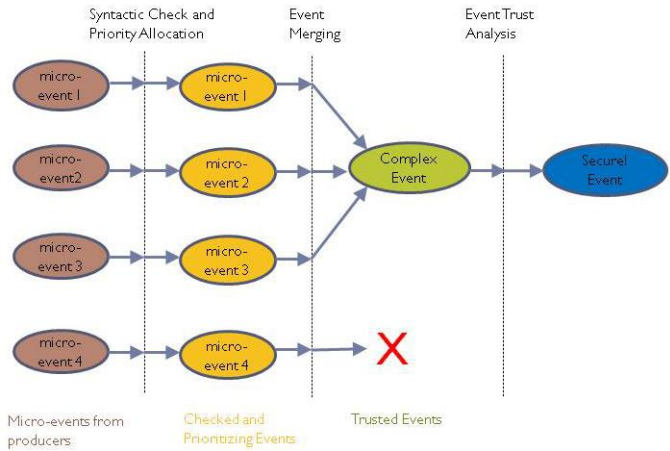

Fig. 3. The correlation process of the micro-events

### B. Event Correlation

The *Event Merging* phase is implemented using the *Esper* engine. Each set of rules that *Esper* can use is stored in the *Rule Repository* of the EPM component and can be activated on-demand by Secure! operators in the operative centers. In order to detect different *complex-event* types belonging to a particular application domain, each rule set characterizes a specific *complex-event* type. This means that each rule set identifies only one *complex-event* type. In this way, a collection of rule sets detects all *complex-event* types in the particular application domain.

*Esper* uses listeners to apply rule sets to the *micro-events*. A listener is an object used to analyze the incoming *micro-events* checking if a *micro-event* set matches with the rules. Each listener contains a specific set of rules and applies continuously the queries defined by them to all *micro-events* incoming from the *Event Bus*. Every time a new *micro-event* reaches the correlator engine, the listeners executes the queries. The listeners can be configured and updated at run-time, this allows a high level of adaptability of the system to the possible changes of its environment (the real context in which Secure! operates).

After the *complex-event* has been built fusing the relevant information from *micro-events*, *event trust analysis* is performed on it in order to produce the *Secure! event*. The latter is sent to the *Situation Extraction and Awareness* level of the Framework for building the overall *Secure! situation* used for decision support. The following Fig. 4 depicts the correlation process based on rules in the *Event processor*.

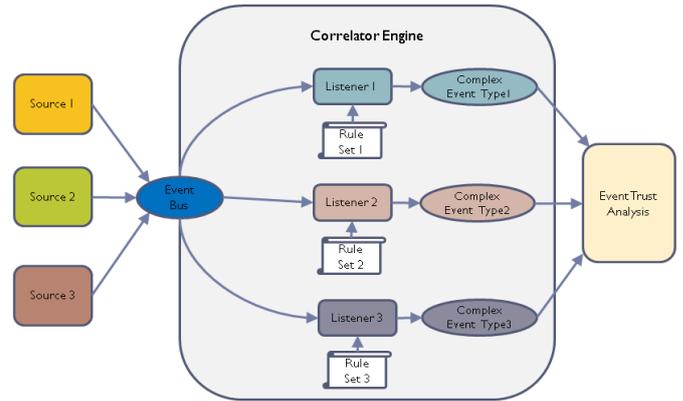

Fig. 4. The correlator engine based on rule sets.

### VII. THE CASE STUDY

The "world-heritage protection" scenario was selected in Secure! as the case study where experimenting the Secure! Framework and hence implementing the EPM component. In this particular context several real events may be dangerous not only for monuments, but also for people security such as: public demonstrations, acts of vandalism (predetermined or not predetermined), armed robberies and so on. When one of these events happens or is going to happen, a set "small" events occurs in the same place at the same time. These events are tweets, phone calls, images, videos, sounds, etc. They are correlated *micro-events* suggesting that a most important and *complex-event* is happening. The world heritage protection scenario is planned to be adopted as a test case of the Secure! project, and specifically regarding the surveillance area of the Miracles Square in Pisa, Italy (a UNESCO World Heritage site, particularly famous for the Leaning Bell Tower).

### A. Implementation of the Case Study

The taxonomy of the *micro-event*, defined in Secure! project, contains a large number of *micro-event* types, for our scope we can consider only a part of them useful to recognize *complex-events* in the considered scenario (or the application domain). Table I lists the *micro-event* types used in the test.

TABLE I. MICRO-EVENT TYPE.

| *Micro-event Type* | *Description* |
|---|---|
| Object Recognition | Recognition of objects belonging to a specific category through the content of video and image analysis. The object categories are those relevant for the public security (shotgun, gun, knife, etc.) |
| People Recognition | Recognition of person identity through the content of video and image analysis |
| People Detection | Detection of people in a specific place. It is relevant if at a certain time or place there is a critical situation in progress. |
| People with Object Recognition | Detection of a person with an object belongings to a relevant category |
| Logo Recognition | Detection of relevant logos on flags, banners, etc. |

| Micro-event Type | Description |
|---|---|
| Trend Detection | Discovery of social communities, or social communication with aggressive or violent purposes (attack organizations, sabotages, acts of vandalism, etc.) |
| Suspicious Spech Recognition | Recognition of keywords in file audio, for example: bomb, weapon, drug and so on. |
| Anomaly Detection | Detection of a potentially dangerous objects near possible target such as monuments |

The *micro-event* types described represent relevant information for the considered application domain. In the performed test, the EPM component receives in input relevant *micro-events*, that have to be correlated, mixed with *noise*. *Noise* is a stream of *micro-events* that the correlation engine should not correlate in the application domain defined by rules. Some set of relevant *micro-events* have been sent in input to EPM component mixed with *noise* in order to estimate the output. The output is considered correct if it comprises only the *complex-events* formed by correlating the relevant *micro-event* according to the rules.

To perform this test a set of EPL rules has been defined for detecting the *complex-events* in the considered scenario. The set of used rules and their corresponding *complex-event* type detected are shown in the table below.

TABLE II. COMPLEX-EVENT TYPE.

| Complex-Event Type | Micro-events | EPL rule |
|---|---|---|
| Dangerous Object positioned by a person | Anomaly Detection, Object Recognition, People Detection | select * from pattern [ every (event1 = Event (anomaly("bag") or anomaly ("knapsack")) and event2=Event ((personBehaviour ("suspicious") )) where timer:within(60 min)] where (event1.timeDiff (event2) <30 and event1.distanceGPS (event2)<0.1) |
| Act of Vandalism | People With Object Detection, Suspicious Human Behaviour Recognition | select * from pattern [ every ( event1 =Event (object ("Hummer") or object ("bar")) and event2 =Event((personBehaviour ("suspicious"))) where timer:within (60 min)] where (event1.timeDiff(event2)<20 and event1.distanceGPS (event2)<0.5) |
| Armed Robbery | People With Object Detection, Suspicious Human Behaviour | select * from pattern [ every ( event1=Event(personBehaviour ("suspicius") and (object ("gun") or object ("knife")) and event2 = Event(audio ("wallet") or audio ("money") or audio ("bag")) where timer:within (5 min)))] where (event1.distanceGPS(event2)<0.4 and event1.timeDiff (event2)<20) |
| Melee | Anomaly Detection, Suspicious Crowd Behaviour Detection | select * from pattern [ every ( event1=Event(people>10) and event2=Event (object ("bar") or object ("knife"))) where timer:within(9 min)] where (event1.distanceGPS(event2) <0.4 and event1.timeDiff (event2)<20) |
| Demonstration | Object Recognition, People Detection | select * from pattern [ every ( event1=Event(people>80) and event2=Event(object ("banner") or object ("svastika") or object ("sickle and hummer"))) where timer:within(9 min)] where (event1.distanceGPS(event2)<0.4 and event1.timeDiff(event2)<20) |

The first column of Table II contains a short description of the *complex-event* types recognized by the detection of a particular set of *micro-events* listed in the second column. The third column shows the EPL correlation rules used to correlate the *micro-events* depending on time and spatial conditions (presently we do not consider the potential localization unaccuracy of devices [19]). The causal correlation is achieved specifying the keywords that define the relationships among the *micro-events*. The used rules are simple examples but useful for our scope, they are extensible to detect more elaborated *complex-events*.

### B. Running the test

The *micro-events* in input to the EPM component are produced automatically by a *micro-event* software generator. They are not real events but they represent realistic events usable for the scope of the test. This choice is due to the fact that the main components of Secure! Framework are not yet integrated because the Framework is actually in development phase. The tests were executed on an Intel G645T CPU 2.5GHz with 4Gb of RAM.

### C. Performance Evaluation

Performance measures have been evaluated during the test activity. We evaluated the mean delay between (i) the arrival of the last *micro-event* belonging to a set that have to be fused in a *complex-event*, and (ii) the generation of the corresponding *complex-event*. We evaluated the mean processing time (mean delay) at varying of *noise* levels. *Noise* is measured by the number of noisy *micro-events* per second, while mean processing time is measured by milliseconds. Fig. 5 depicts the results of the performance test of EPM component.

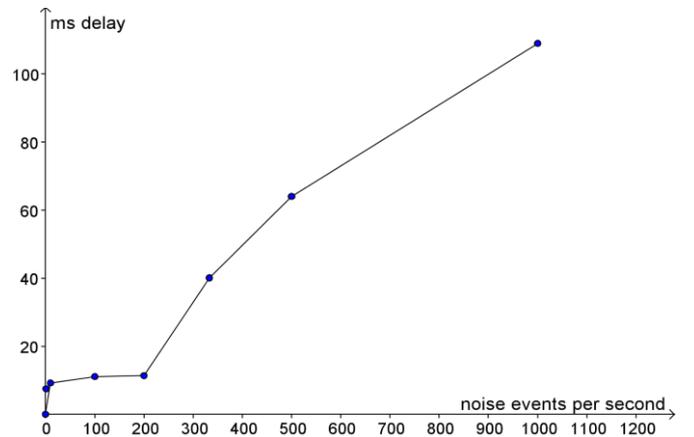

Fig. 5. The mean time of EPM processing time on varying of noise

The graph shows that for low values of *noise*, the EPM component is not affected by relevant delay and *complex-events* are detected almost immediately. As *noise* grows up, delay grows up approximately in linear way.

The test demonstrates that the developed EPM component recognizes *complex-events* starting from a set of *micro-events* even if disturbed by *noise*. In addition, the evaluated mean delay in the correlation process of the EPM component demonstrates that *Esper* is an adequate correlation engine even for real DSS applications.

*D. Accuracy evaluation*

A functional test was performed running the EPM component many times varying *noise* rate in order to evaluate the accuracy of the process. The *micro-events* in input were processed without the presence of errors in output (*complex-events* not detected or false positives) and the 100% of the expected correlations were exactly produced. This shows that *noise* does not affect the correlation in the EPM component.

VIII. CONCLUSIONS AND FUTURE WORK

This work have proposed an approach that uses event correlation technologies to achieve critical situation detection as a support for taking decisions (DSS); event correlation was implemented by exploiting the on-line analysis of real events using the CEP and crowd sensing/sourcing technologies.

Some experiments has been performed in the context of the case study aiming to quantitatively evaluate measures about correct functionality and performances.

Collected results have demonstrated the approach is general (easy to instance, easy to maintain) and extensible to other scenarios where the application requires near real-time correlation, like intrusion detection system [20] and monitoring system of critical infrastructures (e.g. Smart Grid).

Future evolutions of this work will be the assessment of the EPM component using *micro-events* collected in a complex and realistic scenario, namely the world heritage protection scenario described before, and the extension of the measures related with the evaluation of the correlation engine. In addition, due to its extensible nature, future directions will be also to apply this approach to other scenarios, in particular to apply it in Smart Grid infrastructures, where the monitoring of both grid and network domain is not yet well explored [22].

ACKNOWLEDGMENTS

This work was partially supported by Regione Toscana through the "Secure!" Research project (POR-CREO 2007-2013), the PRIN Project "TENACE-Protecting National Critical Infrastructures from Cyber Threats" (n. 20103P34XC) funded by the Italian Ministry of Education, University and Research, and the European Project SmartC2Net (grant agreement no 318023).